\documentclass[%
 reprint,
superscriptaddress,
 amsmath, amssymb,aps,
floatfix,
]{revtex4-1}

\usepackage{graphicx}
\usepackage{dcolumn}
\usepackage{bm}
\usepackage[colorlinks=true,citecolor=blue,linkcolor=magenta]{hyperref}
\usepackage[utf8]{inputenc}
\usepackage{url}
\usepackage[normalem]{ulem}
\usepackage{upgreek}
\usepackage[mathlines]{lineno}
\usepackage{amsmath}

\begin{document}

\title{III-Nitride nanophotonics for beyond-octave soliton generation and self-referencing}

\author{Xianwen Liu}
\affiliation{Department of Electrical Engineering, Yale University, New Haven, CT 06511, USA}

\author{Zheng Gong}
\affiliation{Department of Electrical Engineering, Yale University, New Haven, CT 06511, USA}

\author{Alexander W. Bruch}
\affiliation{Department of Electrical Engineering, Yale University, New Haven, CT 06511, USA}

\author{Joshua B. Surya}
\affiliation{Department of Electrical Engineering, Yale University, New Haven, CT 06511, USA}

\author{Juanjuan Lu}
\affiliation{Department of Electrical Engineering, Yale University, New Haven, CT 06511, USA}

\author{Hong X. Tang}
\affiliation{Department of Electrical Engineering, Yale University, New Haven, CT 06511, USA}
\affiliation{Corresponding author: hong.tang@yale.edu}
\date{\today}
\renewcommand{\figurename}{Fig.}

\begin{abstract}
\vspace{12pt}
Frequency microcombs, successors to mode-locked laser and fiber combs, enable miniature rulers of light for applications including precision metrology, molecular fingerprinting and exoplanet discoveries. 
To enable frequency ruling functions, microcombs must be stabilized by locking their carrier-envelop offset frequency. So far, the microcomb stabilization remains compounded by the elaborate optics external to the chip, thus evading its scaling benefit. To address this challenge, here we demonstrate a nanophotonic chip solution based on aluminum nitride thin films, which simultaneously offer optical Kerr nonlinearity for generating octave soliton combs and Pockels nonlinearity for enabling heterodyne detection of the offset frequency. The agile dispersion control of crystalline III-Nitride photonics permits high-fidelity generation of solitons with features including 1.5-octave spectral span, dual dispersive waves and sub-terahertz repetition rates down to 220\,gigahertz. These attractive characteristics, aided by on-chip phase-matched aluminum nitride waveguides, allow the full determination of the offset frequency. Our proof-of-principle demonstration represents an important milestone towards fully-integrated self-locked microcombs for portable optical atomic clocks and frequency synthesizers. 
\end{abstract}

\maketitle
\section{Introduction}

Optical frequency combs, originally developed from solid-state or fiber based mode-locked lasers, have evolved into photonic-chip-based sources that are compact, robust and power efficient \cite{Fortier201920}. Among various chipscale schemes \cite{Waldburger2019Tightly,Zhang2019Broadband,Gaeta2019Photonic,Pasquazi2018Micro-combs}, microresonator Kerr frequency combs (``microcombs'' hereafter) are of particular interest because of their high scalability for photonic integration \cite{Kippenberg2018Dissipative, Stern2018Battery,Raja2019Electrically}. Indeed, substantial efforts have been made towards soliton mode-locking, allowing phase coherent microcombs on the one hand \cite{Herr2013Temporal,Xue2015Mode-locked,Brasch2015Photonic,yi2015soliton,Joshi2016Thermally,Yu2016Mode-locked,Gong2018High-fidelity,He2019Self-starting, Gong2019Soliton} and unveiling rich soliton physics on the other hand \cite{Yang2016Stokes,Guo2016Universal,Cole2017Soliton}. Specifically, octave-spanning soliton microcombs permit phase locking of the carrier-envelop offset (CEO) frequency ($f_\mathrm{ceo}$) via well-known $f$--$2f$ interferometry \cite{Jones2000Carrier}, and are prerequisite for chip-scale implementation of precision metrology \cite{Giunta201920}, frequency synthesizers \cite{Spencer2018An} and optical clocks \cite{Newman2019Architecture}. To date, silicon nitride (Si$_3$N$_4$) nanophotonics has proved viable for octave soliton operations with a terahertz repetition rate ($f_\mathrm{rep}$) \cite{Li2017Stably,Pfeiffer2017Octave,Yu2019Tuning}. Nevertheless, such a large $f_\mathrm{rep}$ is not amenable for direct photodetection and poses challenges to access the CEO frequency with a value up to $f_\mathrm{rep}$. In the meantime, the lack of intrinsic quadratic $\chi^{(2)}$ nonlinearities in Si$_3$N$_4$ films typically requires an external frequency doubler and off-chip optical circuitry for deriving the CEO frequency \cite{Drake2019Terahertz,Briles2018Interlocking}. These off-chip optical components compromise the scaling advantage of microcombs and significantly set back self-locked microcombs for portable applications.

III-Nitride semiconductors such as aluminum nitride (AlN) exhibit a non-centrosymmetric crystal structure, thereby possessing inherent optical $\chi^{(2)}$ nonlinearity as well as Pockels electro-optic and piezoelectric properties \cite{xiong2012aluminum}. Apart from the advances in ultraviolet light-emitting diodes \cite{kneissl2019emergence} and quantum emitters \cite{bishop2020room,lu2020bright}, AlN has also proved viable for low-loss nanophotonics in high-efficiency second-harmonic generation (SHG) \cite{Bruch201817000,Liu2019Beyond} and high-fidelity Kerr and Pockels soliton mode-locking \cite{Gong2018High-fidelity,Bruch2020Pockels}. Therefore, it is feasible to establish an on-chip $f$--2$f$ interferometer provided that an octave AlN soliton microcomb is available. This is a solution that is favored here comparing with the heterogeneous integration approach such as proposal based on hybrid gallium arsenide (GaAs)/Si$_3$N$_4$ waveguides \cite{chang2018heterogeneously}. 
Despite that on-chip $f_\mathrm{ceo}$ detection was achieved from supercontinuua driven by a femtosecond laser in non-resonant $\chi^{(2)}$ nanophotonic waveguides made from AlN \cite{Hickstein2017Ultrabroadband} or lithium niobate (LN) thin films \cite{yu2019coherent, okawachi2020chip}, resonator microcomb-based $f$--$2f$ interferometry using nanophotonics, to our knowledge, remains elusive. 

In this article, we demonstrate high-fidelity generation of octave soliton microcombs and subsequent $f_\mathrm{ceo}$ detection using AlN-based nanophotoinc chips. Thanks to mature epitaxial growth, AlN thin films with highly uniform thickness are available, thus permiting lithographic control of group velocity dispersion (GVD) for comb spectral extension via dispersive wave (DW) emissions \cite{Brasch2015Photonic}. Our octave soliton microcombs possess separated dual DWs and moderate $f_\mathrm{rep}$ of 433, 360 and 220 gigahertz, and are found to be reproducible from batch-to-batch fabrications. The results then allow us to capture the $f$--$2f$ beatnote through on-chip SHG in phase-matched AlN waveguides. Our work establish the great potential of non-centrosymmetric III-Nitride photonic platforms for portable self-locked microcomb sources.

\begin{figure*}[!t]
\centering
\includegraphics[width=\linewidth]{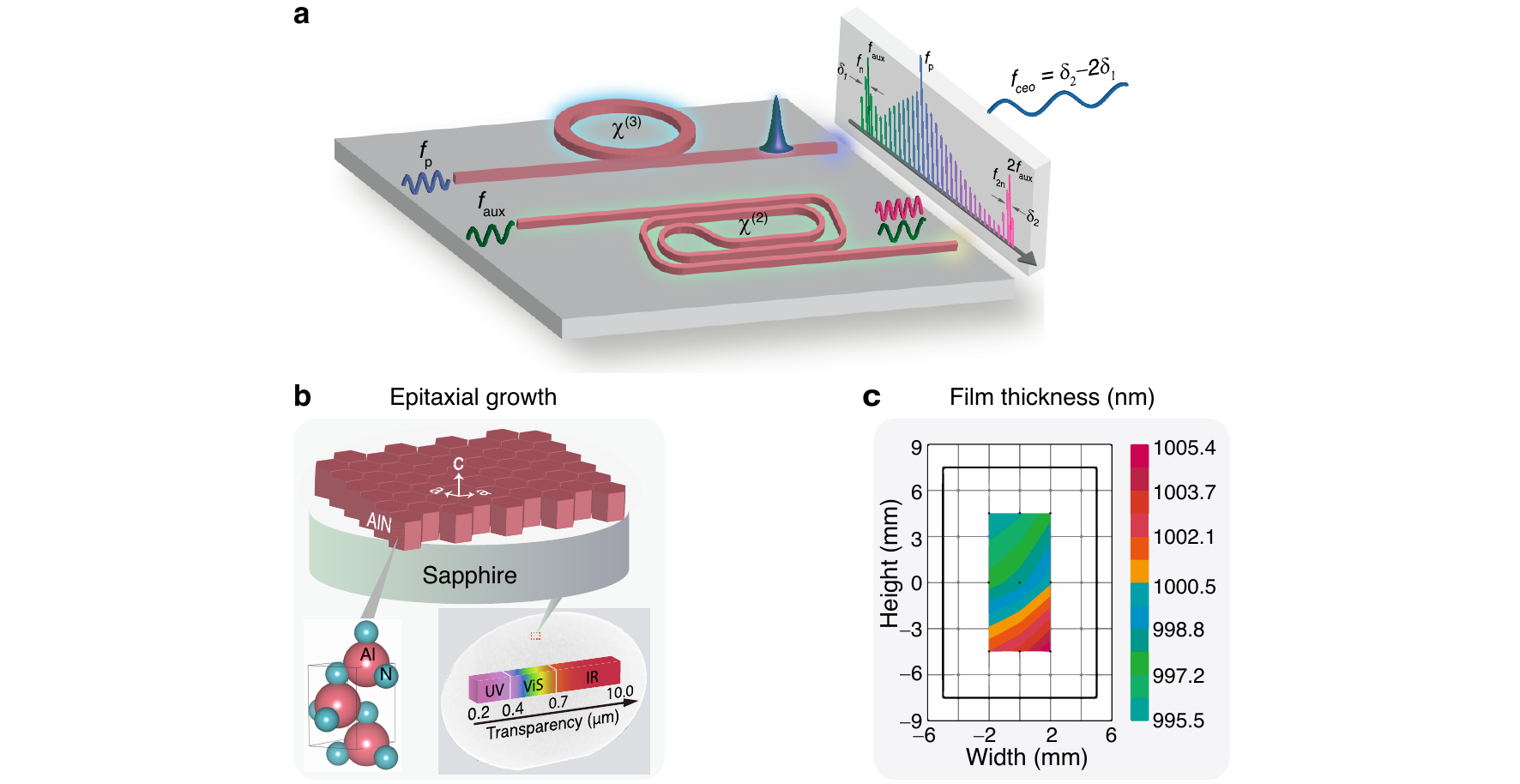}
\caption{\textbf{Experimental scheme}. \textbf{a}. Illustration of $f$--$2f$ interferometry using octave-spanning soliton microcombs and second-harmonic generators in a nanophotonic platform harboring simultaneous $\chi^{(3)}$ and $\chi^{(2)}$ susceptibilities. The offset frequency $f_\mathrm{ceo}$ is accessible from the beatnotes of $\delta_1$ and $\delta_2$, and $f_\mathrm{p}$ is the pump laser frequency. \textbf{b}. Top: sketch of a hexagonal AlN layer (lattice constants: $a$ and $c$) epitaxially grown on a $c$-plane sapphire substrate. Bottom: unit cell of an AlN crystal (left) and photograph image of a 2-inch AlN wafer featuring a broad transparency window from ultraviolet to mid-infrared regimes (right). \textbf{c}. Spectroscopic ellisometer mapping of the AlN film thickness in a region of 4\,$\times$\,9\,mm$^2$, showing a minor variation of $1000\pm5$\,nm denoted by the right color bar.}
\label{fig1}
\end{figure*}

\section{Results}
\noindent \textbf {Experimental scheme description.} Figure\,\ref{fig1}a illustrates the implementation of microcomb-based $f$--2$f$ interferometry using nanophotonic chips. The strategy is to leverage non-centrosymmetric photonic media for simultaneous integration of $\chi^{(3)}$ octave soliton microcombs and $\chi^{(2)}$ SHG doublers. For a proof-of-principle demonstration, we adopt an auxiliary laser (at $f_\mathrm{aux}$) to obtain sufficient SHG power (at 2$f_\mathrm{aux}$) from phase-matched optical waveguides. The use of the auxiliary laser can be eliminated by exploiting microring-based architecture to boost the SHG efficiency \cite{Bruch201817000}. By subsequently beating $f_\mathrm{aux}$ and 2$f_\mathrm{aux}$ with the $f_\mathrm{n}$ and $f_\mathrm{2n}$ comb lines at their corresponding beatnotes of $\delta_{1}$ and $\delta_{2}$, the $f_\mathrm{ceo}$ signal reads:
\begin{equation}
f_\mathrm{ceo}=\delta_{2}-2\delta_{1}
\label{eq1}
\end{equation}

The AlN thin films in this work were epitaxially grown on a $c$-plane sapphire substrate via metal-organic chemical vapour deposition \cite{liu2017integrated,liu2018integrated}. As illustrated in Fig.\,\ref{fig1}b, the AlN crystals exhibit a hexagonal wurtzite structure with a unit cell shown in the bottom, highlighting the non-centrosymmetry. We also show an overall 2-inch AlN-on-sapphire wafer featuring a broadband transparency and a favored film thickness (Fig.\,\ref{fig1}c)--both are crucial factors to ensure octave GVD control. Great attention was also paid to the film crystal quality and surface roughness for low-loss photonic applications. The AlN nanophotonic chips were manufactured following electron-beam lithography, chlorine-based dry etching and silicon dioxide (SiO$_{2}$) coating processes and were subsequently cleaved to expose waveguide facets \cite{liu2018ultra}. The intrinsic optical quality factors ($Q_\mathrm{int}$) of the AlN resonators were characterized to be $\sim$1--3 million depending on the waveguide geometries. The detailed film and device characterization is presented in Methods and Supplementary Section\,I.

\begin{figure*}[!t]
\centering
\includegraphics[width=\linewidth]{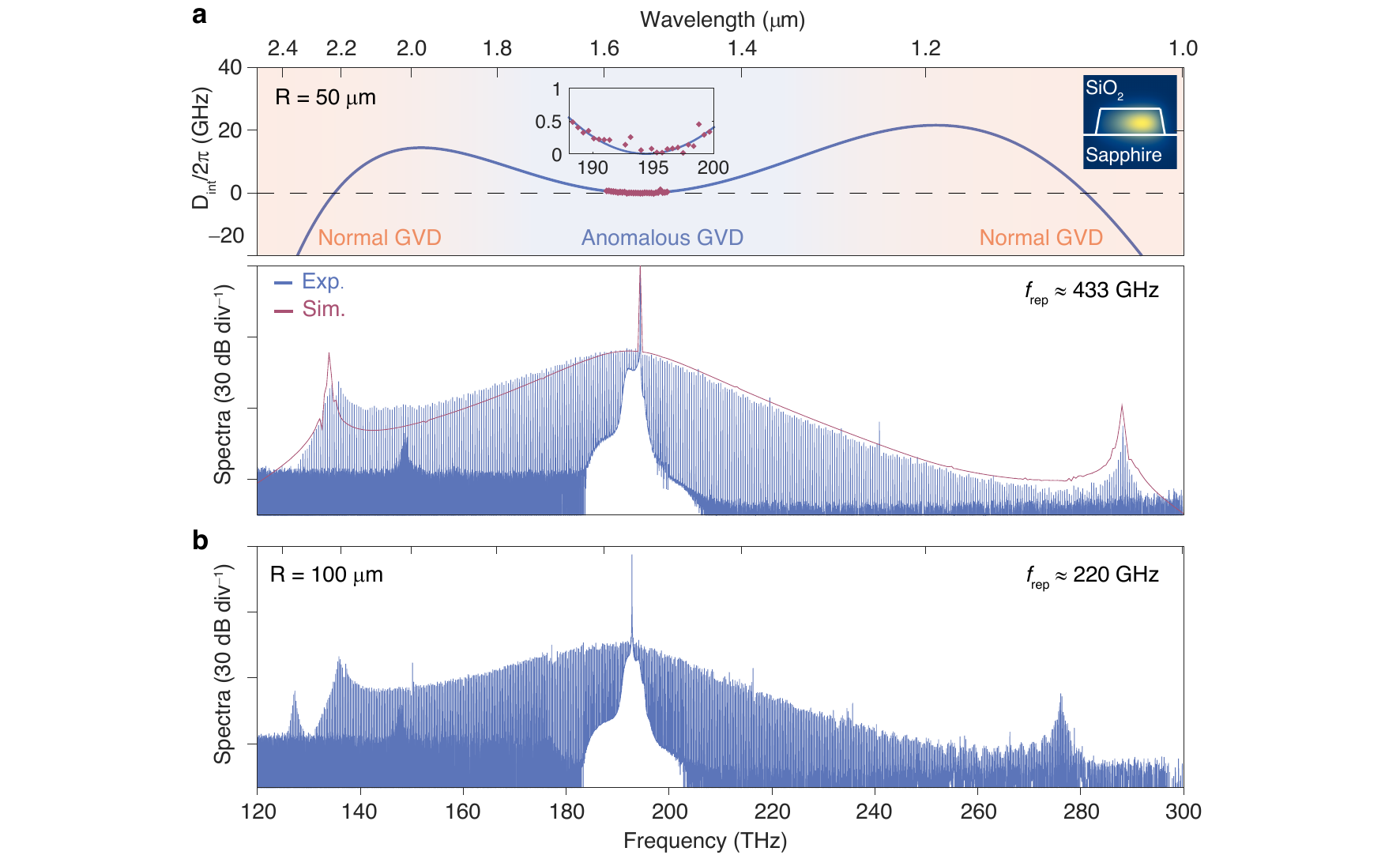}
\caption{\textbf{Octave soliton microcombs at hundreds of gigahertz repetition rates}. \textbf{a}. Top: integrated dispersion ($D_\mathrm{int}$) of a 50\,$\mu$m-radius AlN resonator (cross section: 1.0\,$\times$\,2.3\,$\mu$m$^2$), where the anomalous and normal GVD regimes are shaded with light blue and orange colors respectively. Insets: zoom-in view of the measured (red dots) and simulated (blue curve) values and the resonator modal profile shown in the right. Bottom: soliton microcomb spectra from the experiment (blue) and simulation (red) at an on-chip pump power of $\sim$390\,mW. The resonator $Q_\mathrm{int}$ is 1.6\,million and the $f_\mathrm{rep}$ is estimated to be around 433\,GHz. \textbf{b}. Soliton microcomb spectrum from a 100\,$\mu$m-radius AlN resonator (cross section: 1.0\,$\times$\,3.5\,$\mu$m$^2$) with a decreased $f_\mathrm{rep}$ of $\sim$220\,GHz. The applied pump power is $\sim$1\,W at a resonator $Q_\mathrm{int}$ of 3.0\,million.}
\label{fig2}
\end{figure*}

Since wurtzite AlN manifests optical anisotropy for vertically or horizontally-polarized light \cite{majkic2017optical}, we engineer the waveguide structures for optimal operation of fundamental transverse magnetic (TM$_{00}$) modes, which allows the harness of its largest $\chi^{(2)}$ susceptibility to ensure high-efficiency SHG. To expand microcomb spectra out of the anomalous GVD restriction, we exploit soliton-induced DW radiation by tailoring the resonator's integrated dispersion ($D_\mathrm{int}$) \cite{Brasch2015Photonic}:
\begin{equation}
D_\mathrm{int}=\frac{D_{2}}{2!}\mu^{2}+\frac{D_{3}}{3!}\mu^{3}+\sum_{i\geq4}\frac{D_{i}}{i!}\mu^{i}
\label{eq2}
\end{equation}
where $D_2$, $D_3$, and $D_i$ are $i_\mathrm{th}$-order GVD parameters, while $\mu$ indexes the relative azimuth mode number with respect to the pump ($\mu$\,=\,0).

\noindent \textbf {Octave soliton microcombs.} Our GVD engineered AlN resonators are coated with a SiO$_{2}$ protection layer, making it less susceptible to the ambient compared with the air-cladded Si$_3$N$_4$ counterpart \cite{Li2017Stably,Pfeiffer2017Octave,Briles2018Interlocking,Yu2019Tuning,Drake2019Terahertz}. An example of the resonator modal profile is shown in the inset of Fig.\,\ref{fig2}a. The top panel of Fig.\,\ref{fig2}a plots the $D_\mathrm{int}$ curve from a 50\,$\mu$m-radius AlN resonator through numerical simulation (see Methods).  In spite of the limited anomalous GVD window (light blue shade), octave microcomb operation is feasible via DW radiations at phase-matching conditions $D_\mathrm{int}$\,=\,0, allowing for spectral extension into normal GVD regimes (light orange shade). Note that the occurrence of such dual DWs benefits from the optimal film thickness in our AlN system, while the DW separation is agilely adjustable over one octave through the control of resonator's dimensions. (see Supplementary Section\,II). Around the telecom band, the $D_\mathrm{int}$ value (red dots) was characterized by calibrating the resonator's transmission with a fiber-based Mach–Zehnder interferometer \cite{Gong2018High-fidelity,Gong2019Soliton}. The experimental result matches well with the simulated one (inset of Fig.\,\ref{fig2}a) with an extracted $D_2/2\pi$ of $\sim$6.12\,MHz.

\begin{figure*}[t!]
\centering
\includegraphics[width=\linewidth]{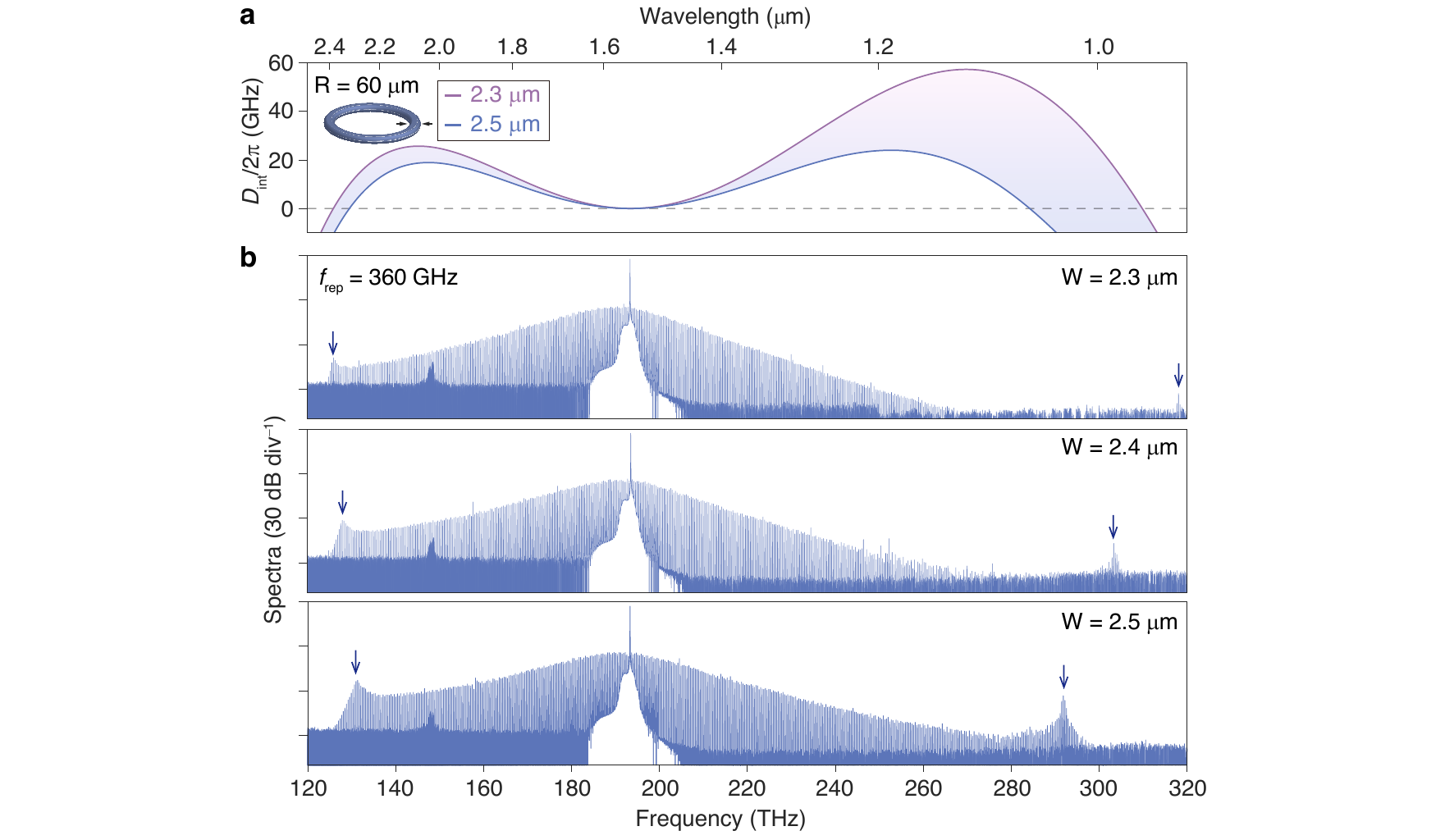}
\caption{\textbf{Octave soliton microcombs with agilely tunable spectra}. \textbf{a}. Engineered $D_\mathrm{int}$ curves of 60\,$\mu$m-radius AlN resonators at varied widths of 2.3--2.5\,$\mu$m revealed by the colored shadow regime. \textbf{b}. Corresponding soliton microcomb spectra at resonator widths of 2.3, 2.4, and 2.5\,$\mu$m from the top to bottom panel, respectively. The  $f_\mathrm{rep}$ is $\sim$360\,GHz, while the vertical arrows in spectral wings indicate the emergence of DWs. Akin to Fig.\,\ref{fig2}a, high-frequency DWs here also exhibit an evident blue shift from the $D_\mathrm{int}$\,=\,0 position. From a sech$^2$ fit, the corresponding temporal pulse duration is estimated to be $\sim$23, 22 and 19\,fs (from top to bottom), respectively.}
\label{fig3}
\end{figure*}

We then explore soliton mode-locking based on a rapid frequency scan scheme to address the abrupt intracavity thermal variation associated with transitions into soliton states \cite{Gong2018High-fidelity}. The soliton spectrum is recorded using two grating-based optical spectrum analyzers (OSAs, coverage of 350--1750\,nm and 1500--3400\,nm). The experimental setup is detailed in Supplementary Section\,II. The bottom panel of Fig.\,\ref{fig2}a plots the soliton spectrum from a 50\,$\mu$m-radius AlN resonator, featuring a moderate $f_\mathrm{rep}$ of 433\,GHz and an observable spectral span of 1.05--2.4\,$\mu$m, exceeding one optical octave. Meanwhile, soliton-induced DW radiations occur at both ends of the spectrum, in agreement with the predicted $D_\mathrm{int}$ curves. Note that the low-frequency DW location matches well with the $D_\mathrm{int}$\,=\,0 position, while the high-frequency one exhibits an evident blue shift, which is ascribed to Raman-induced soliton red shifts relative to the pump frequency \cite{karpov2016raman,gong2020near}. This conclusion is supported by the soliton spectral simulation (red curve) when accounting for Raman effects (see Methods), while the intact low-frequency DW might be a result of the cancellation of soliton recoils \cite{Brasch2015Photonic}.  

The single crystal nature of AlN thin films permits reproducible optical index in each manufacture run. This, in combination with their uniform film thickness control, leads to a high predictability for the dispersion engineering, making it feasible to predict octave soliton combs at various repetition rates. For instance, our GVD model indicates that octave spectra with repetition rates further decreased by two times are anticipated from 100\,$\mu$m-radius AlN resonators at optimal widths of 3.3--3.5\,$\mu$m (see Supplementary Section\,II). Figure\,\ref{fig2}b plots the recorded soliton comb spectrum at a resonator width of 3.5\,$\mu$m, where a $f_\mathrm{rep}$ of $\sim$220\,GHz and dual DWs separated by more than one octave are achieved simultaneously. Such a low $f_\mathrm{rep}$ is amenable for direct photodetection with state-of-the-art unitravelling-carrier photodiodes \cite{zhang2019terahertz}. We also noticed the occurrence of a weak sharp spectrum around 130\,THz, which might arise from modified local GVD due to avoided mode crossing\cite{yi2017single}. In our nanophotonic platform, we could further predict resonator geometries for achieving octave solitons with an electronically detectable $f_\mathrm{rep}$ of $\sim$109\,GHz (see Supplementary Section\,II). Nonetheless, the strong competition between Kerr nonlinearities and stimulated Raman scattering (SRS) must be taken into account since the free spectrum range (FSR) of the resonator is already smaller than the A$_1^\mathrm{TO}$ phonon linewidth ($\sim$138\,GHz) in AlN crystals \cite{liu2018integrated}.

\begin{figure*}[t!]
\centering
\includegraphics[width=\linewidth]{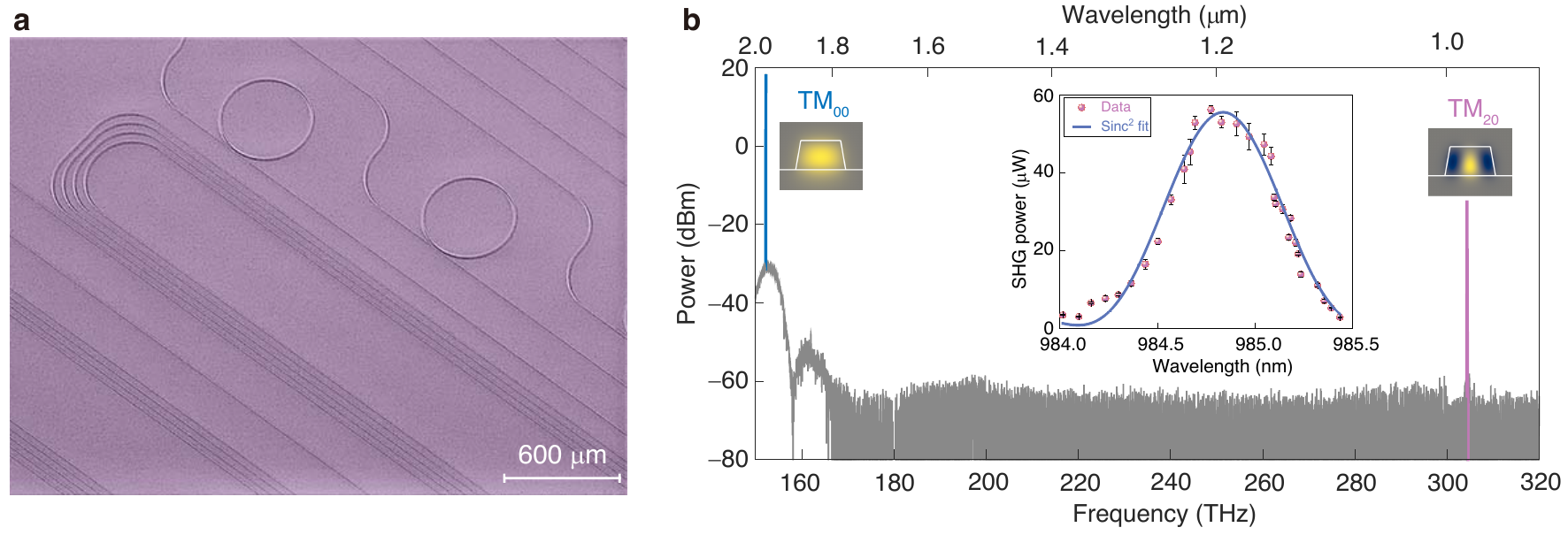}
\caption{\textbf{Second-harmonic generators.} \textbf{a}. Colored scanning electron microscope images of fabricated AlN nanophotonic chips composed of octave microcomb generators (microring resonators) and SHG waveguides (total length of 6\,cm, not fully shown). \textbf{b}. SHG spectra collected from a modal phase-matched waveguide (width of 1.395\,$\mu$m) at an on-chip $1f$ power of 355\,mW. Insets: modal profiles of the pump (TM$_{00}$) and SHG (TM$_{20}$) waves as well as the wavelength-dependent SHG power (pink dots), where a sinc$^2$-function fit (blue curve) is applied. The error bars reflect the SHG power variation from continuous three measurements.
}
\label{fig4}
\end{figure*}

Since the SHG from the auxiliary laser (1940--2000\,nm) available in our laboratory is beyond the soliton spectral coverage shown in Fig.\,\ref{fig2}, we further adjust the resonator dimensions for extending microcomb spectra below 1\,$\mu$m. As plotted in Fig.\,\ref{fig3}a, the phase-matching condition ($D_\mathrm{int}$\,=\,0) for high-frequency DW radiations below 1\,$\mu$m is fulfilled by elevating the resonator radius to 60\,$\mu$m while maintaining its width around 2.3\,$\mu$m. In the meantime, low-frequency DWs could also be expected and their spectral separation is adjustable by controlling the resonator width. Guided by the tailored $D_\mathrm{int}$ curves, we fabricated the AlN resonators and recorded octave soliton spectra at a $f_\mathrm{rep}$ of $\sim$360\,GHz (Fig.\,\ref{fig3}b). Lithographic control of DW radiations (indicated by vertical arrows) is also verified by solely adjusting the resonator width, allowing the spectral extension below 1\,$\mu$m (width of 2.3 or 2.4\,$\mu$m). The low- and high-frequency DWs are found to exhibit distinct frequency shifting rates, consistent with the $D_\mathrm{int}$ prediction. The observable soliton spectra (from top to bottom of Fig.\,\ref{fig3}b) cover 1.5, 1.3, and 1.2 optical octaves by normalizing the total span ($\Delta f$) to its beginning frequency ($f_1$), that is $\Delta f/f_1$. Such a definition permits a fair comparison among soliton microcomb generation in distinct pump regimes across different material platforms, suggesting high competitiveness of our AlN microcomb span comparing to state-of-the-art values reported in Si$_3$N$_4$ microresonators \cite{Pfeiffer2017Octave}.

\noindent \textbf{On-chip second harmonic generator.} 
We then explore the co-integration of SHG based on the $\chi^2$ susceptibility of AlN for matching the DW peak below 1\,$\mu$m (middle panel of Fig.\,\ref{fig3}b). To fulfill the demanding requirement of spectral overlaps with the microcomb, we adopt a straight waveguide configuration, which allows a broader phase-matching condition albeit at the cost of reduced conversion efficiencies comparing to its counterpart using dual-resonant microresonators \cite{Bruch201817000,Bruch2019On-chip}.
Through modeling, we predict an optimal waveguide width of $\sim$1.38\,$\mu$m for fulfilling the modal-phase-matching condition (see Supplementary Section\,III), while the actual waveguide width was lithographically stepped from 1.32 to 1.46\,$\mu$m (spacing of 5\,nm) accounting for possible deviations during the manufacturing process. 

Figure\,\ref{fig4}a shows a section of 6\,cm-long SHG waveguides co-fabricated with the microcomb generator. At a fixed fundamental wavelength (1970\,nm), we located the phase-matching waveguide at the width of 1.395\,$\mu$m, close to the predicted width. The corresponding SHG spectra are plotted in Fig.\,\ref{fig4}b, where we achieve a high off-chip SHG power over 50\,$\mu$W by boosting the fundamental pump power from a thulium-doped fiber amplifier to compensate the SHG efficiency (see Supplementary Section\,III). In the meantime, the wavelength-dependent SHG power shown in the inset indicates a large 3-dB phase-matching bandwidth of $\sim$0.8\,nm, which, together with an external heater for thermal fine-tuning, is sufficient to cover the target comb lines for subsequent heterodyne beating. 

\begin{figure}[t!]
\centering
\includegraphics[width=\linewidth]{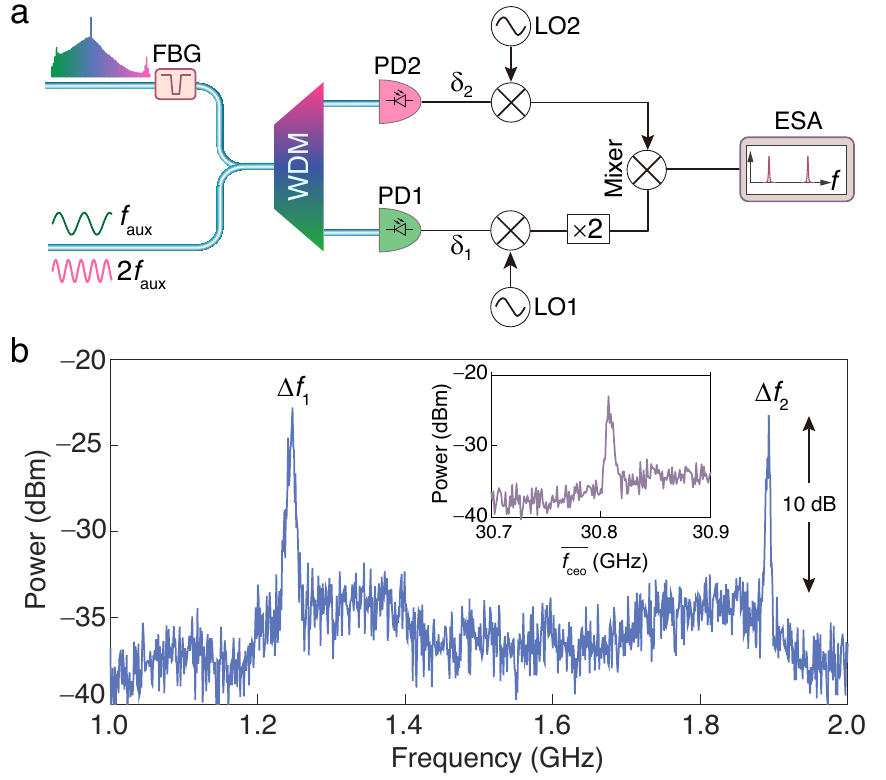}
\caption{\textbf{$f$--$2f$ heterodyne measurement.} \textbf{a}. Schematic diagram for assessing the $f_\mathrm{ceo}$. The symbol ``$\times2$'' indicates a RF frequency doubler. The experimental details are introduced in Supplementary Section\,III. \textbf{b}. Free-running $f$--$2f$ beatnotes after the down-conversion process, suggesting a signal-to-noise ratio of 10\,dB at a resolution bandwidth of 1\,MHz. The local oscillator frequencies $f_\mathrm{LO1}$ and $f_\mathrm{LO2}$ are chosen to be 11.8 and 9.1\,GHz, respectively. Inset: the equivalent curve of $\overline{f_\mathrm{ceo}}$\,=\,$2f_\mathrm{LO1}$\,+\,$f_\mathrm{LO2}$\,--\,$\Delta f_2$.}
\label{fig5}
\end{figure}

\noindent \textbf{Nanophotonic $f$--$2f$ interferometry.} 
By combining outgoing light from optimal AlN soliton and SHG generators on the calibrated OSAs, we are able to estimate the $f$--$2f$ beatnote frequency to be approximately 32\,GHz limited by the resolution of the OSAs. To electronically access the $f_\mathrm{ceo}$ signal in real time, we employ a scheme sketched in Fig.\,\ref{fig5}a. The recorded soliton spectrum after suppressing pump light by a fiber Bragg grating (FBG) indicates a high off-chip power close to $-40$\,dBm for the high-frequency DW (see Supplementary Section\,III). Meanwhile, a wavelength-division multiplexer (WMD) is utilized to separate the $f$ and $2f$ frequency components before sent into the photodetectors (PDs). Two tunable radio frequency (RF) synthesizers are introduced as the local oscillators (LO1 and LO2) to down convert the photodetector signals for effective capture of $f$--$2f$ beat signal at a convenient low-frequency band with an electronic spectrum analyzer (ESA, range of 20\,Hz--26.5\,GHz). 

As highlighted in Fig.\,\ref{fig5}b, we record two down-converted beatnotes of $\Delta f_1$ and $\Delta f_2$ with a signal-to-noise ratio of 10\,dB at a resolution bandwidth of 1\,MHz. Much higher signal-to-noise ratios are anticipated by applying a finer detection bandwidth upon locking the telecom pump laser as well as the $f_\mathrm{ceo}$ frequency \cite{Spencer2018An, Newman2019Architecture}. The corresponding $f$--$2f$ beatnote is $\overline{f_\mathrm{ceo}}$\,=\,$2f_\mathrm{LO1}$\,+\,$f_\mathrm{LO2}$\,--\,$\Delta f_2$ (inset of Fig. \ref{fig5}b) since the local oscillator frequencies $f_\mathrm{LO1}$ and $f_\mathrm{LO2}$ are chosen to be larger than beatnotes of $\delta_1$ and $\delta_2$. We also note that the actual $f_\mathrm{ceo}$ in the current device equals to $\mathrm{FSR}$\,--\,$\overline{f_\mathrm{ceo}}$, which is unveiled by tuning the relative positions of auxiliary laser and comb teeth frequencies, indicating the involvement of $f_\mathrm{n}$ and $f_\mathrm{2n+1}$ comb lines in the heterodyne beating (see Supplementary Section\,III). On the other hand, the $f_\mathrm{LO1}$ and $f_\mathrm{LO2}$ frequencies are freely adjustable up to 40 and 20\,GHz in our scheme, which could further expand the accessible range of $f_\mathrm{ceo}$ frequency based on the down-conversion process presented here. Meanwhile, the RF synthesizers are synchronized to a common external frequency reference, suggesting that the captured down-converted $f$--$2f$ signals are available for further locking the comb teeth in a feedback loop as reported in Ref. \cite{Drake2019Terahertz}. 

\section{Discussion}
We demonstrate nanophotonic implementation of $f$--$2f$ interferometry by leveraging $\chi^{(3)}$ octave solitons and $\chi^{(2)}$ SHG co-fabricated from a non-centrosymmetric AlN photonic platform. Thanks to agile GVD engineering offered by epitaxial AlN thin films, our octave soliton microcombs can reliably produce dual DWs and sub-THz repetition rates (220--433\,GHz) that are accessible with unitravelling-carrier photodiodes. The overall soliton spectral span is adjustable up to 1.5 octave, on a par with state-of-the-art values (1.4 octave) reported in Si$_3$N$_4$ microresonators. We further perform the $f_\mathrm{ceo}$ measurement with the aid of an auxiliary laser for enabling SHG in phase-matched AlN waveguides, thus allowing for spectral overlap with the desired octave soliton.

For future development, the spectral restriction of octave solitons for matching with the auxiliary laser wavelength can be relaxed by exploiting high-efficiency SHG in dual-resonant microresonators, which allows direct doubling of a selected comb line in the low frequency DW band \cite{surya2018control}. Meanwhile, the octave comb's repetition rate can be further reduced by leveraging on-chip Pockels electro-optical frequency division \cite{li2014electro}. By shifting the phase matching condition for SHG, it is also possible to extend octave solitons into the near-visible band, giving access to self-locked near-visible microcombs for precision metrology. Besides, the exploration of other non-centrosymmetric photonic media such as LN, GaAs and gallium phosphide could be envisioned \cite{gong2020near,chang2018heterogeneously, Wilson2019Integrated}. For instance, by exploiting periodically poled LN thin films \cite{Lu2019Periodically}, it is likely to simultaneously achieve phase matched $\chi^{(2)}$ and octave $\chi^{(3)}$ interactions in a single microring resonator, thus simplifying the photonic architectures. Our results represent an important milestone to unlock the potentials of octave microcomb technologies for portable applications.

\section{Methods}
\noindent \textbf{Nanofabrication.} 
The surface roughness and crystal quality of our AlN thin films were respectively characterized by an atomic force microscope and an X-ray diffraction scan, indicating a root-mean-square roughness of 0.2\,nm in $1\times1$\,$\mu$m$^2$ region and an FWHM linewidth of $\sim$46 and 1000\,arcsec along (002) and (102) crystal orientations, respectively. The film thickness was mapped by a spectroscopic ellipsometer (J.A. Woollam M-2000), providing a quick and preliminary selection of the desired AlN piece for octave soliton generation with dual DWs. As shown in Supplementary Section\,I, in spite of varied film thicknesses across a 2-inch AlN wafer, we can reliably locate the desired region for reproducible octave device fabrication.  

To further reduce the propagation loss, the AlN photonic chips were annealed at 1000\,$^{\circ}$C for 2\,hours. The resonator Q-factors were probed by sweeping a tunable laser (Santec TSL-710) across the cavity resonances and then fitted by a Lorentzian function. In the 100\,$\mu$m-radius AlN resonators (width of 3.5\,$\mu$m), we achieve a recorded Q$_\mathrm{int}$ of 3.0\,million, while the 50\,$\mu$m-radius resonators (width of 2.3\,$\mu$m) exhibit a decreased Q$_\mathrm{int}$ of 1.6\,million, indicating the dominant sidewall scattering loss of our current fabrication technology. The related resonance curves are plotted in Supplementary Section\,I.

\noindent \textbf{Numerical simulation.} 
The $D_\mathrm{int}$ of the AlN resonators is investigated using a finite element method (FEM) by simultaneously accounting for the material and geometric chromatic dispersion. The overall $D_\mathrm{int}$ value is approximated with a fifth-order polynomial fit applied to the simulated modal angular frequencies: $\omega_\mu$\,=\,$\omega_0$\,+\,$\mu D_1$\,+\,$D_\mathrm{int}$, where $D_1/(2\pi)$ is the resonator's FSR at the pump mode $\mu$\,=\,0.

The spectral dynamics of octave soliton microcombs is numerically explored based on nonlinear coupled mode equations by incorporating the Raman effect \cite{gong2020near,gong2020photonic}:
\begin{multline}
\frac{\partial}{\partial t}a_{\mu}=-(\frac{\kappa_\mu}{2}+i\Delta^a_\mu)a_{\mu}+ig_\mathrm{K}\sum_{k,l,n}a_{k}^{*}a_{l}a_{n}\delta(l+n-k-\mu)\\-ig_\mathrm{R}\sum_{k,l}a_{l}\big[\mathcal{R}_{k}\delta(l+k-\mu)+\mathcal{R}_{k}^{*}\delta(l-k-\mu)\big]+\xi_{\mathrm{P}}
\label{eq3}
\end{multline} 

\begin{multline}
\frac{\partial }{\partial t}\mathcal{R}_{\mu}=-(\frac{\gamma_\mathrm{R}}{2}+i\Delta^\mathrm{R}_\mu)\mathcal{R}_{\mu}-ig_\mathrm{R}\sum_{k,l}a_{k}^{*}a_{l}\delta(l-k-\mu)
\label{eq4}
\end{multline} 
Here $a$ and $\mathcal{R}$ are the mode amplitudes of cavity photons and Raman phonons with subscripts $k,l,n$ being the mode indices, while $g_\mathrm{K}$ and $g_\mathrm{R}$ represent the nonlinear coupling strength of Kerr and Raman processes, respectively. The driving signal strength is $\xi_{\mathrm{P}}=\delta(\mu)\sqrt{\frac{\kappa_{\mathrm{e},0}P_{\mathrm{in}}}{\hbar\omega_{\mathrm{p}}}}$ at an on-chip pump power $P_\mathrm{in}$, $\kappa_{\mu}$ ($\kappa_{\mathrm{e},\mu}$) denotes the total (external) cavity decay rate of the $\mu^{\mathrm{th}}$ photon mode, and $\gamma_{\mathrm{R}}$ is the Raman phonon decay rate. The detuning from a $D_1$-spaced frequency grid is indicated by $\Delta^a_\mu$\,=\,$\omega_\mu$\,--\,$\omega_\mathrm{P}$\,--\,$\mu{D_{1}}$ and $\Delta^\mathrm{R}_\mu$\,=\,$\omega_{\mathrm{R}}$\,--$\mu{D_{1}}$ with $\omega_\mathrm{P}$ and $\omega_\mathrm{R}$ being pump and Raman shift angular frequencies, respectively.

In the simulation, we set the time derivative of Raman items in Eq.\,\ref{eq4} to zero to speed up the computation since the decay rate of phonons is much larger than that of photons. We also consider frequency-independent $\kappa_\mu/(2\pi)$\,$\approx$\,120\,MHz and $\kappa_{\mathrm{e},\mu}/(2\pi)$\,$\approx$\,75\,MHz based on measured Q-factors of 50\,$\mu$m-radius AlN resonators in Fig.\,\ref{fig2}a. Because incident light is TM-polarized, the involved A$_1^\mathrm{TO}$ Raman phonon in AlN exhibits an $\omega_\mathrm{R}/(2\pi)$\,$\approx$\,18.3\,THz with an FWHM of $\gamma_\mathrm{R}/(2\pi)$\,$\approx$\,138\,GHz \cite{liu2017integrated}. The $g_\mathrm{K}/2\pi$ is calculated to be 0.73\,Hz for a given nonlinear refractive index $n_2$\,=\,2.3\,$\times$\,10$^{-19}$\,m$^2$/W, while an optimal $g_\mathrm{R}/2\pi$\,=\,0.29\,MHz is adopted, resulting in a soliton spectrum matching well with the measured one in Fig.\,\ref{fig2}a. The simulated high frequency DW also exhibits an evident blue shift comparing with the case of $g_\mathrm{R}/2\pi$\,=\,0\,MHz (see Supplementary Section\,II).

\section{Data availability} The data that support the findings of this study are available from the corresponding authors upon reasonable request.

\def\bibsection{\section{\textbf{references}}}

\bibliographystyle{myaipnum4-1}
\bibliography{octave_comb}

\vspace{1 mm}
\noindent \textbf{Acknowledgements.} This work was supported by DARPA SCOUT (W31P4Q-15-1-0006). H.X.T. acknowledges partial support from DARPA's ACES program as part of the Draper-NIST collaboration (HR0011-16-C-0118) and DARPA's A-Phi program with a subcontract from Sandia labs (DENA0003525). The authors thank Y.  Sun,  S.  Rinehart and K.  Woods in Yale cleanrooms and  M.  Rooks in YINQE for assistance in the device fabrication, and J. Xie and M. Xu in the laboratory for the microwave circuit discussion.

\vspace{1 mm}
\noindent \textbf{Author contributions.} H.X.T and X.L conceived the idea. X.L. performed the device design, fabrication and measurement with the assistance from Z.G., A.B., J.S. and J.L.. Z.G. and X.L. performed the soliton simulation. X.L. and H.X.T. wrote the manuscript with the input from all other authors. H.X.T supervised the project.

\vspace{2 mm}
\noindent \textbf{Competing interests.} The authors declare no competing interests.

\vspace{2 mm}
\noindent \textbf{Additional information.} Supplementary information accompanies this manuscript.

\end{document}


\rmfamily
\Large
\textbf{Supplementary Material:}

\vskip 10pt
\large
\title {\bf{III-Nitride nanophotonics for beyond-octave soliton generation and self-referencing}}

\author{Xianwen Liu$^1$, Zheng Gong$^1$, Alexander W. Bruch$^1$, Joshua B. Surya$^1$, Juanjuan Lu$^1$, and Hong X. Tang}
\normalsize
\vskip 5pt

\affiliation{Department of Electrical Engineering, Yale University, New Haven, CT 06511, USA}

\affiliation{Corresponding author: hong.tang@yale.edu}
\maketitle

\setcounter{figure}{0} 
\renewcommand{\thefigure}{\textbf{S\arabic{figure}}}
\renewcommand{\figurename}{\textbf{Fig.}}
\renewcommand{\theequation}{S\arabic{equation}}
\renewcommand{\bibnumfmt}[1]{[S#1]}
\renewcommand{\citenumfont}[1]{S#1}

\setcounter{equation}{0} \renewcommand{\theequation}{S\arabic{equation}}
\sffamily
\large

\normalsize
\vskip 10pt
\nopagebreak

\section{Device fabrication}
\begin{figure*}[!h]
\centering
\includegraphics[width=\linewidth]{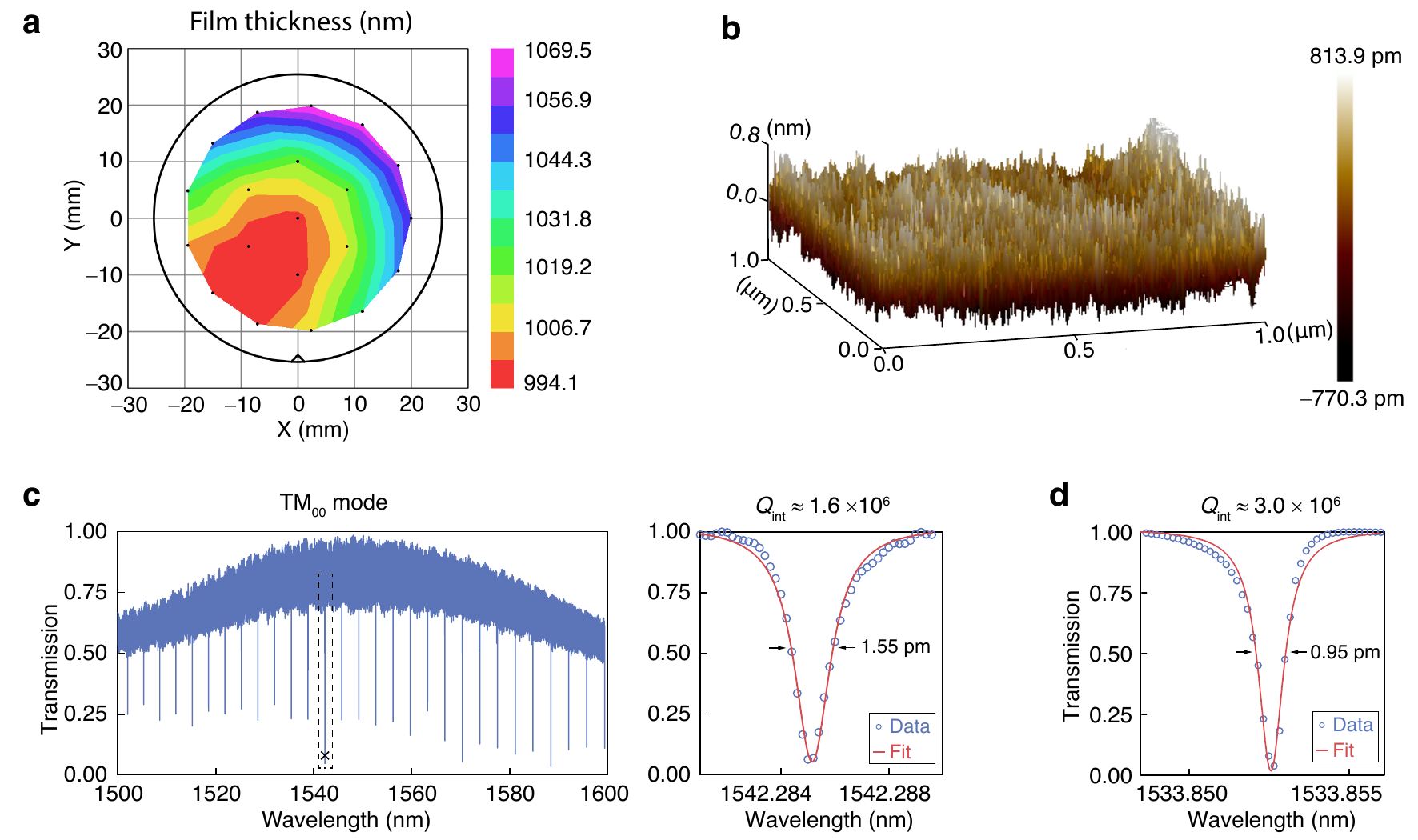}
\caption{\textbf{Nanophotonic device fabrication}. \textbf{a}. Wafer-scale thickness mapping of a 2-inch AlN wafer using a spectroscopic ellipsometer. The right color bar indicates the corresponding thickness variation. \textbf{b}. Surface roughness characterization of the AlN film using an atomic force microscope. \textbf{c}. Transmittance of a 50\,$\mu$m-radius AlN resonator (width\,=\,2.3\,$\mu$m) for octave-soliton generation. A zoom-in view of the resonance (indicated by dashed lines) around 1542\,nm is shown in the right side, revealing a $Q_\mathrm{int}$ of $\sim$1.6 million. \textbf{d}. Resonance from an 100\,$\mu$m-radius AlN resonator (width\,=\,3.5\,$\mu$m), highlighting a reduced resonant linewidth and an improved $Q_\mathrm{int}$ of $\sim$3.0 million.}
\label{figS1}
\end{figure*}

All devices are patterned from 2-inch crystalline AlN-on-sapphire wafers with 1000\,nm-thick AlN epilayers. To ensure robust dispersion engineering for reproducible octave-soliton generation, we first characterized the uniformity of film thickness across the wafer. As shown in Fig.\,\ref{figS1}a, the film thickness of the wafer is very close to target growth thickness apart from the edge of the wafer (colored by purple). By intentionally choosing the desired thickness region, octave-soliton generation can be reliably realized using our current nanofabrication technology. For low-loss nanophotonic applications, we also paid attention to the crystal quality of the AlN thin film, whose root-mean-square surface roughness was characterized to be as low as 0.2\,nm in a 1\,$\times$\,1\,$\mu$m$^2$ region. The result is presented in Fig.\,\ref{figS1}b.

An example of the resonator transmittance is shown in Fig.\,\ref{figS1}c, where the fundamental transverse magnetic (TM$_{00}$) mode is effectively excited while higher-order modes are  suppressed by adopting a pulley waveguide coupling configuration (wrapped angle of 6$^{\circ}$). The intrinsic quality-factors ($Q_\mathrm{int}$) are then extracted from a Lorentz fit of the resonance curve at under-coupled conditions. The AlN resonators engineered for octave-soliton generation exhibits a dimension-dependent $Q_\mathrm{int}$ of 1.6 million and 3.0 million for the devices with radii of 50\,$\mu$m (width\,=\,2.3\,$\mu$m, right of Fig.\,\ref{figS1}c) and 100\,$\mu$m (width\,=\,3.5\,$\mu$m, Fig.\,\ref{figS1}d), suggesting the dominant sidewall scattering loss. Further improvement of the $Q$-factors can be envisioned by leveraging the racetrack resonators, where the straight portions exhibit a much smoother sidewall \cite{zhang2017Monolithic}.  

\vskip 20pt
\large
\section{Octave Soliton Engineering and Characterization}
\begin{figure*}[!h]
\centering
\includegraphics[width=\linewidth]{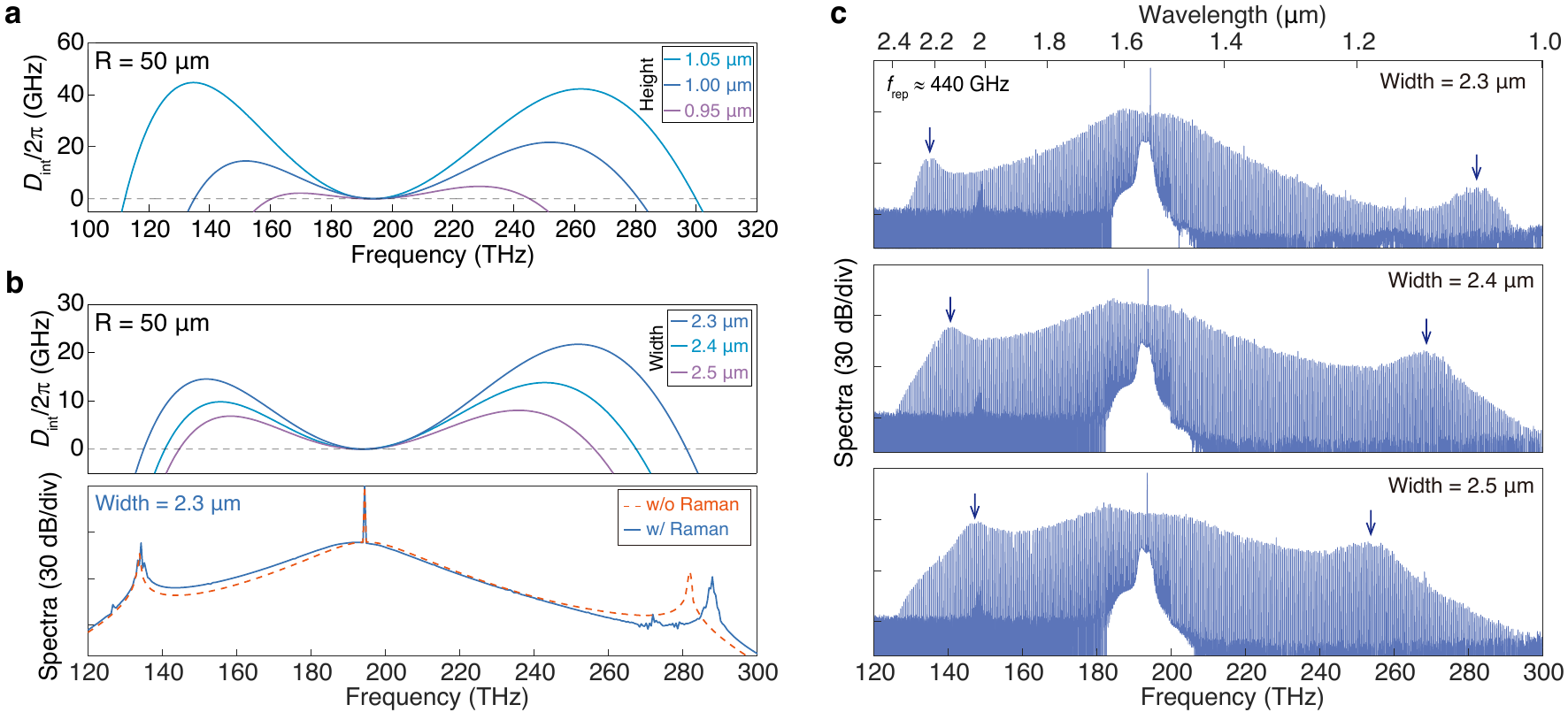}
\caption{\textbf{Dispersion engineering for octave soliton generation}. \textbf{a}. Height-dependent integrated dispersion ($D_\mathrm{int}$) of 50\,$\mu$m-radius AlN resonators at a fixed width of 2.3\,$\mu$m. \textbf{b}. Top: width-dependent $D_\mathrm{int}$ curves (radius of 50\,$\mu$m, height of 1.00\,$\mu$m). Bottom: numerically simulated soliton comb spectra without (orange) or with (blue) the influence of Raman effects for the $D_\mathrm{int}$ curve at a resonator width of 2.3\,$\mu$m (height of 1.0\,$\mu$m). \textbf{c}. Experimentally recorded chaotic comb spectra at a varied resonator width of 2.3, 2.4 and 2.5\,$\mu$m. The vertical arrows indicate the emergence of dispersive wave-like envelopes, and the corresponding $D_\mathrm{int}$ curves are shown in the top panel of \textbf{b}.}
\label{figS2}
\end{figure*}

\normalsize
We note that the AlN resonator's integrated dispersion ($D_\mathrm{int}$) is highly susceptible to the film thickness variation, which affects octave soliton generation with phase-matched dual dispersive waves (DWs). As plotted in Fig.\,\ref{figS2}a, when the resonator height deviates from an optimal value of 1.00\,$\mu$m (blue curve), such as increasing to 1.05\,$\mu$m or decreasing to 0.95\,$\mu$m, a larger dispersion barrier or a narrower dispersion window will occur on both sides of the $D_\mathrm{int}$ curve, preventing from octave spectral extension via DW radiations. As a result, we intentionally locate the AlN piece with the desired thickness around 1000\,nm (see Fig.\,\ref{figS1}a) for the octave-soliton device fabrication. 

At an optimal height of 1.0\,$\mu$m, the $D_\mathrm{int}$ curve can be further engineered by tailoring the resonator width. The result is shown in the top panel of Fig.\,\ref{figS2}b, where the phase-matching condition ($D_\mathrm{int}$\,=\,0) for DW radiations is readily adjusted beyond one octave span when reducing the resonator width from 2.5 to 2.3\,$\mu$m. We then numerically investigate the octave soliton spectrum for the $D_\mathrm{int}$ curve at a width of 2.3\,$\mu$m. As shown in the bottom panel of Fig.\,\ref{figS2}b, the high-frequency DW exhibits an evident blue shift from the $D_\mathrm{int}$\,=\,0 position when accounting for the Raman effect, which matches wells with our experimental result in Fig.\,2a of the main text. The underlying mechanism for this spectral shift is attributed to the Raman-induced soliton red shift in the spectral center, which in turn blue shifts the high-frequency DW \cite{Gong2020near}. Figure\,\ref{figS2}c plots the noise-state comb spectra recorded from the dispersion engineered AlN resonators (radius of 50\,$\mu$m, width of 2.3--2.5\,$\mu$m). It is found that the DW-like envelops at both ends of the spectra exhibit an evident shift when varying the resonator width, in good agreement with the $D_\mathrm{int}$ curve prediction (top panel of Fig.\,\ref{figS2}b). In our experiment, the corresponding soliton spectrum can be captured at a resonator width of 2.3\,$\mu$m (see Fig.\,2a in the main text), while it is inaccessible at the width of 2.4 and 2.5\,$\mu$m due to the occurrence of Raman lines in the intermediate state, thus hampering soliton mode-locking.

\begin{figure*}[!h]
\centering
\includegraphics[width=\linewidth]{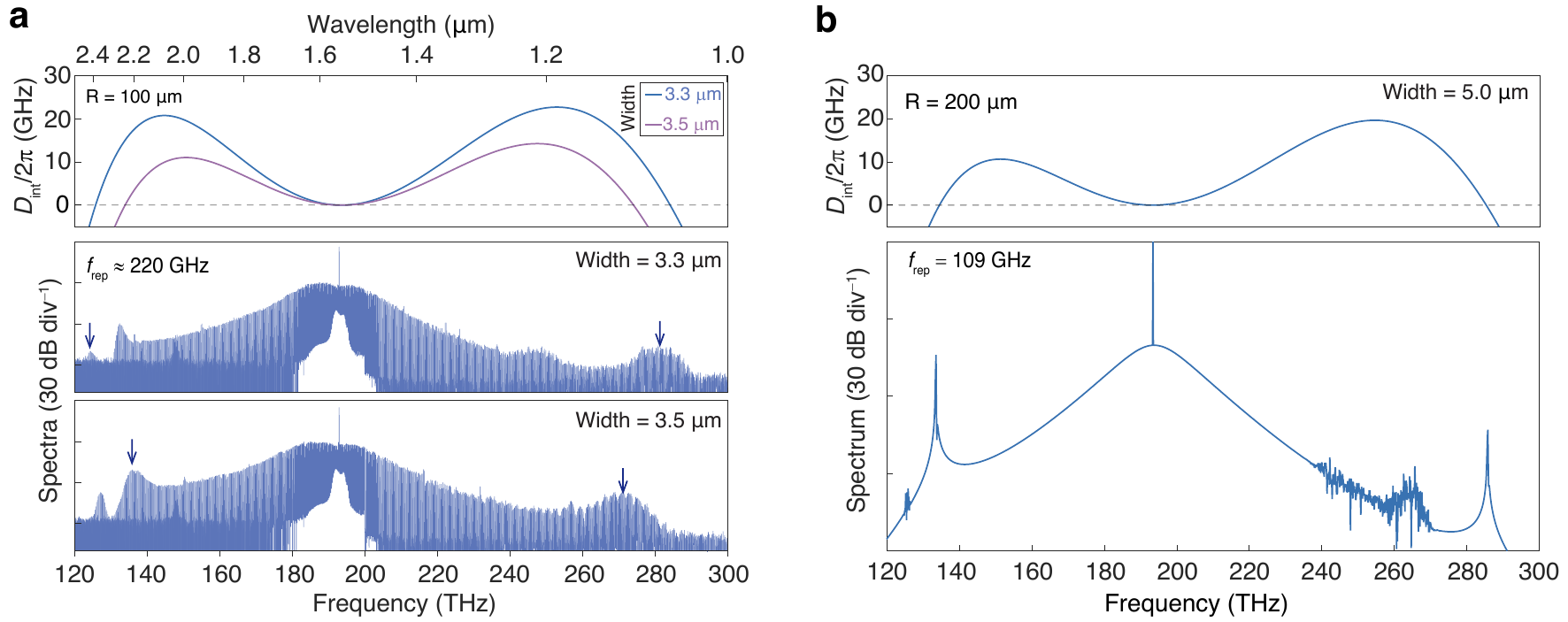}
\caption{\textbf{Dispersion engineering for repetition rate-detectable octave solitons}. \textbf{a}. Top: $D_\mathrm{int}$ curves of engineered AlN resonators at a radius of 100\,$\mu$m and varied widths of 3.3 and 3.5\,$\mu$m. Bottom: captured noise-state comb spectra with a reduced repetition rate of $\sim$220\,GHz. The vertical arrows indicate the emergence of DW-like peaks. \textbf{b}. Top: $D_\mathrm{int}$ curve of AlN resonators at an increasing radius of 200\,$\mu$m and an optimal width of 5.0\,$\mu$m. Bottom: simulated soliton spectrum at an on-chip pump power of 100\,mW, highlighting a reduced repetition rate of 109\,GHz.}
\label{figS3}
\end{figure*}

By further engineering the resonator dimensions, we are able to achieve octave solitons with repetition rates reduced by two times. As shown in the top panel of Fig.\,\ref{figS3}a, at an elevated resonator radius of 100\,$\mu$m, the phase-matching
conditions for separated dual DWs by one optical octave are accessible from engineered $D_\mathrm{int}$ curves at optimal resonator widths of 3.3 and 3.5\,$\mu$m. This prediction is verified by the recorded comb spectra with repetition rates ($f_\mathrm{rep}$) of $\sim$220\,GHz (bottom panel of Fig.\,\ref{figS3}a), where dual DW-like envelopes indicated by vertical arrows match well with the $D_\mathrm{int}$\,=\,0 condition in each case. The abnormal spectral peaks observed in the low-frequency region might arise from the avoided mode-crossing due to imperfections in the device fabrication \cite{Herr2014mode}, which is not included in our dispersion modeling. The soliton comb spectrum was captured at a resonator width of 3.5\,$\mu$m (see Fig.\,2b of the main text), while the occurrence of intermediate Raman lines prevents from soliton mode-locking for the case of a resonator width\,=\,3.3\,$\mu$m.

Note that the agile dispersion engineering in our material system also offers the capability to achieve octave solitons with electronically detectable repetition rates by commercial high-speed photodetectors (bandwidth\,\textgreater\,100\,GHz). As shown in Fig.\,\ref{figS3}b, upon increasing the resonator radius to 200\,$\mu$m, it is possible to achieve a low free spectrum range (FSR) of 109\,GHz and an optimal $D_\mathrm{int}$ curve (top panel) for octave soliton generation at a resonator width of 5.0\,$\mu$m. The corresponding soliton spectrum was numerically investigated and presented in the bottom panel. Here we choose an aspirational $Q_\mathrm{int}$ of 10 million at critical coupled conditions for enabling octave soliton generation at a low on-chip pump power of 100\,mW. Since the resonator's FSR is already smaller than the A$_1^\mathrm{TO}$ phonon linewidth ($\sim$138\,GHz) of crystal AlN films \cite{Liu2017integrated}, the Raman effect has to be suppressed for soliton mode-locking as discussed in Ref. \cite{Gong2020near}. For the numerical investigation shown here, we ignore the influence of Raman effects, while it should be considered in the practical devices.

\begin{figure*}[!h]
\centering
\includegraphics[width=\linewidth]{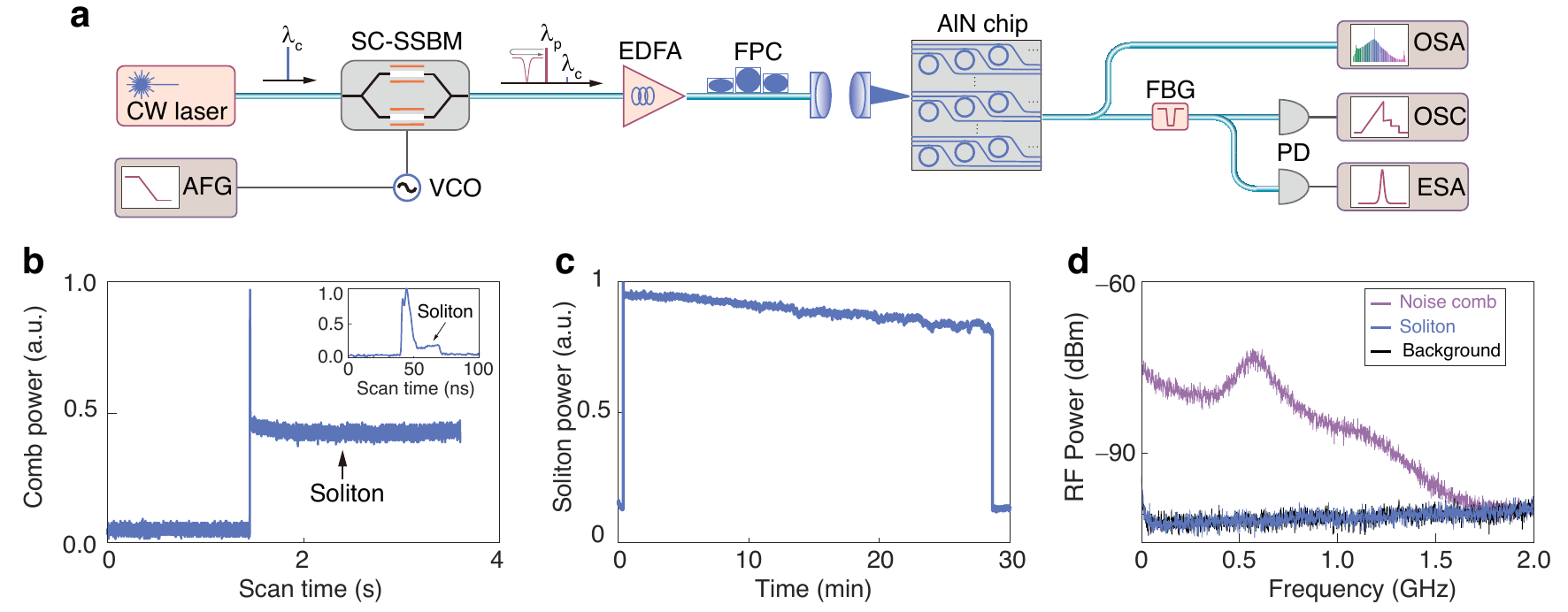}
\caption{\textbf{ Octave-soliton characterization}. \textbf{a}. Sketch of the experimental configuration. The SC-SSBM produces a blue-shifted sideband ($\lambda_p$) relative to incident light ($\lambda_c$) from a continuous-wave (CW) laser. The sideband is then boosted by an erbium-doped fiber amplifier (EDFA) and aligned to the vertical polarization before entering the AlN chip via an aspherical lens pair. \textbf{b}. Comb power trace recorded in the OSC, indicating elongated soliton duration time beyond 2\,s. Inset: initial soliton lifetime of $\sim$20\,ns. \textbf{c}. Free-running soliton power trace as a function of the time. \textbf{d}. Radio-frequency (RF) beating properties of chaotic (red) and soliton microcombs (blue) comparing with the PD background (black).}
\label{figS4}
\end{figure*}

The experimental setup for our octave soliton generation is sketched in Fig.\,\ref{figS4}a. The transition from chaotic to soliton states typically accompanies a notable intracavity power drop, which in turn renders thermo cooling of the resonator with blue-shifted resonances, hindering stable soliton formation \cite{Herr2014temporal}. We address this obstacle using rapid frequency scan schemes based on a suppressed-carrier single sideband modulator (SC-SSBM) \cite{Gong2018high}. The SC-SSBM is driven by a voltage controller oscillator (VCO) connected to an arbitrary function generator (AFG), allowing rapid frequency shifting (up to 500\,MHz/ns) across the resonance at a timescale far beyond the thermal-optic response (microseconds). 

For characterization, light existing the chip is collected by a bare fiber (mode diameter of 4\,$\mu$m) before sent into two grating-based optical spectrum analyzers (OSAs, the other one is not shown) and two photodetectors (PDs) following by an oscilloscope (OSC) and an electronic spectrum analyzer (ESA). A fiber-Bragg grating is also employed to suppress strong pump light. Figure\,\ref{figS4}b plots a typical comb power trace when entering the soliton state. In spite of the initial soliton lifetime of $\sim$20\,ns, we can elongate it beyond 2\,s using the rapid frequency scan scheme. The corresponding octave-soliton spectrum is shown in Fig.\,2a of the main text. Upon entering the soliton state, our octave comb maintains a high stability during the full experiment span until the fiber-to-chip coupling becomes misaligned as indicated in in Fig.\,\ref{figS4}c. We also evaluate the coherence of the spectrum by sending a portion of comb lines (after suppressing pump light) into a PD. As shown in Fig.\,\ref{figS4}d, there is no evidence of low-frequency radio-frequency (RF) noise within a span of 2\,GHz for the octave soliton comparing with the chaotic state, suggesting a high degree of coherence.  

\large
\section{On-chip second harmonic generation and self-referencing}
\normalsize

\begin{figure*}[!htb]
\centering
\includegraphics[width=\linewidth]{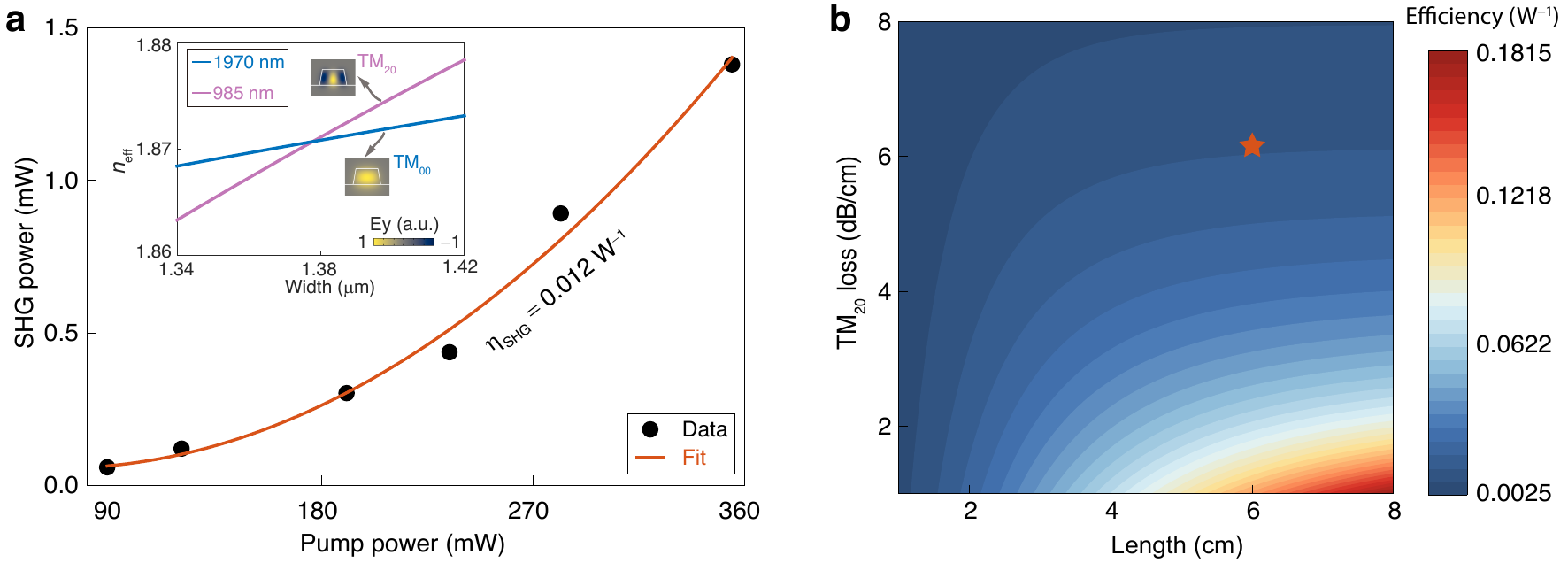}
\caption{\textbf{SHG conversion efficiency}. \textbf{a}. On-chip SHG power versus the pump power as well as the applied second-order polynomial fit (orange solid line). Inset: calculated effective indices ($n_\mathrm{eff}$) of TM$_{00}$ (wavelength of 1970\,nm) and second-order TM$_{20}$ (wavelength of 985\,nm) modes versus the waveguide width. \textbf{b}. Calculated SHG efficiency (indicated by right color bar) as a function of the TM$_{20}$ mode propagation loss and the waveguide length. The "red star" symbol corresponds to our experimentally estimated SHG efficiency.}
\label{figS5}
\end{figure*}

To access the carrier-envelop offset frequency ($f_\mathrm{ceo}$) of octave soliton combs, a set of AlN waveguides were co-fabricated for efficient second-harmonic generation (SHG) as described in Fig.\,4a of the main text. Here we consider the modal-phase-matching case for the pump (TM$_{00}$) and SHG (TM$_{20}$) modes, for which we can obtain an optimal waveguide width around 1.38\,$\mu$m to fulfill the phase-matching condition (inset of Fig.\,\ref{figS5}a). By lithographically scanning the waveguide width at a spacing of 5\,nm, we locate the phase-matching waveguide for producing SHG spectra shown in Fig.\,4b of the main text. The insertion loss of the TM$_{00}$ mode at 1970\,nm was measured to be $\sim$7\,dB/facet, while the insertion loss of the TM$_{20}$ mode at 985\,nm is estimated to be around 15\,dB/facet because of the small modal overlap between the high-order TM$_{20}$ mode with the fiber mode. The calibrated on-chip SHG power ($P_\mathrm{SHG}$) versus the pump power ($P_\mathrm{p}$) is plotted in Fig.\,\ref{figS5}a, where an on-chip SHG conversion efficiency $\eta_\mathrm{SHG}$\,=\,$P_\mathrm{SHG}/P_\mathrm{p}^2$ is derived to be 0.012\,W$^{-1}$.

We then analytically investigate the SHG efficiency with the coupled wave equation at a slowly varying amplitude approximation. Since the SHG mode (TM$_{20}$) is more susceptible to the waveguide sidewall roughness, we also include its propagation loss ($\alpha$) into the model shown below (see supplementary of Ref. \cite{Liu2018generation}):
\begin{equation}
\frac{db}{dz}=-\frac{\alpha}{2}b+\frac{{i\omega_b}^{2}\chi^{(2)}{\Gamma}}{4k_{b}c^2}a^2\mathrm{exp}(i\Delta kz)
\label{equ1}
\end{equation}
Here $a$ and $b$ are the slowly varying field amplitude of TM$_{00}$ and TM$_{20}$ waves along the waveguide direction $z$, while $k_q$\,=\,$n_b\omega_q/c$ is the propagation constant with $n$, $\omega$, and $c$ being the effective index, angular frequency, and light speed in vacuum, respectively. The subscript $q$ denotes the parameters of the mode $a$ or $b$. Meanwhile, $\chi^{(2)}$ is the quadratic optical nonlinearity, $\Delta k$ equals to $2k_a$\,--\,$k_b$, and $\Gamma$ describes the modal overlap of $a$ and $b$ modes given by:
\begin{equation}
\Gamma =\frac{\int u_a^2u_{b}^{*}dxdy}{(\int \left | u_a \right |^{2}dxdy)(\int \left | u_b \right |^{2}dxdy)^{1/2}}
\label{equ2}
\end{equation}
where $u_a$ and $u_b$ indicate the transverse field distribution. 

By solving Eq.\,(\ref{equ1}) for deriving the SHG power $P_\mathrm{SHG}$\,=\,$\frac{n_b\varepsilon_0 c\left | b \right |^{2}}{2}\int \left | u_b \right |^{2}dxdy$, the on-chip SHG efficiency reads:
\begin{equation}
\eta_\mathrm{SHG}=\frac{(\omega_a\chi^{(2)}\Gamma L)^{2}}{2n_a^2n_b\varepsilon_0 c^3}\frac{\mathrm{sinh}^2(\alpha L/4)+\mathrm{sin}^2(\Delta kL/2)}{(\alpha L/4)^2+(\Delta kL/2)^2}
\label{equ3}
\end{equation}
Here $\varepsilon_0$ is the permittivity in vacuum and $L$ is the overall waveguide length. Based on Eq.\,(\ref{equ3}), we calculate $\eta_\mathrm{SHG}$ at phase-matching condition (i.e., $\Delta k$\,=\,0) as shown in Fig.\,\ref{figS5}b. It is evident that the mode propagation loss makes a singficant impact on the SHG efficiency in such a long waveguide. In the experiment, we adopt a waveguide length of 6\,cm and the calculated $\eta_\mathrm{SHG}$ is found to agree with the experimental value when the TM$_{20}$ loss is around 6\,dB/cm. The result is in reasonable agreement with our experimentally extracted TM$_{20}$ loss ($\sim$2\,dB/cm at 780\,nm \cite{Bruch201817000}) when accounting for possible deviation from perfect phase-matching conditions in practical devices.

\begin{figure*}[!htb]
\centering
\includegraphics[width=\linewidth]{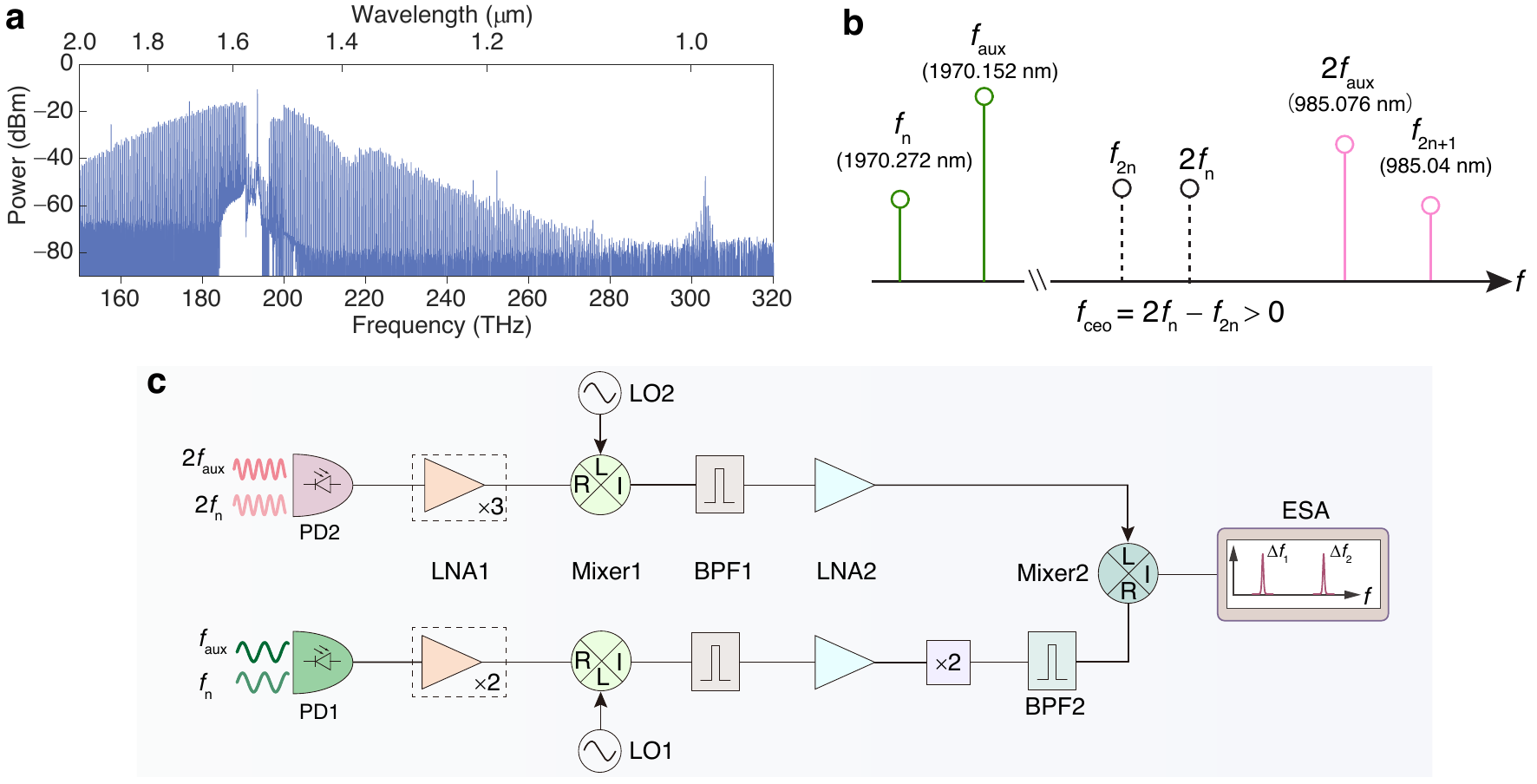}
\caption{\textbf{Implementation of $f$--$2f$ interferometry}. \textbf{a} Octave soliton spectrum after suppressing the pump by a FBG (see full spectrum in Fig.\,4b of the main text). \textbf{b}. Sketch of the involved frequencies (monitored by the OSAs) around the $f$ and $2f$ bands for the $f_\mathrm{ceo}$ measurement. \textbf{c}. Illustration of the experimental setup for electronically accessing the $f_\mathrm{ceo}$ based on a down-conversion frequency process. PD1: 830--2150\,nm, 0--12.5\,GHz; PD2: 800--1700\,nm, 0--26\,GHz; LNA1: 6--18\,GHz, gain\,$\approx$\,27\,dB; LNA2: 10--800\,MHz, gain\,$\approx$\,60\,dB; Mixer1: L/R port (7.5--20\,GHz), I port (0--7.5\,GHz); BPF1: 20--1000\,MHz; RF doubler: 10--1000\,MHz; BPF2: 1.5--2\,GHz; Mixer2: L/R port (1--2700\,MHz), I port (1--2000\,MHz).}
\label{figS6}
\end{figure*}

Based on the optimized octave soliton comb and SHG generator, we establish a $f$--$2f$ interferometer for accessing the $f_\mathrm{ceo}$ frequency. To enable efficient optical-to-electrical conversion in the PDs, the residual pump light is suppressed by a broadband FBG and the recorded octave comb spectrum is presented in Fig.\,\ref{figS6}a. By adjusting the auxiliary laser frequency ($f_\mathrm{aux}$) to overlap with one of the comb line, we produce a strong $2f_\mathrm{aux}$ tone in the proximity of $f_\mathrm{2n}$ comb line. In our case, $2f_\mathrm{aux}$ is actually closer to the $f_\mathrm{2n+1}$ comb line, which is then selected for implementing the $f$--$2f$ interferometry. These optical frequencies are monitored by two OSAs (350--1750\,nm and 1500--3400\,nm) at a resolution of 0.05 and 0.1\,nm, respectively. The positions of relevant laser lines are sketched in Fig.\,\ref{figS6}b, where a positive $f_\mathrm{ceo}$ frequency is ensured at the assigned comb line indices. By setting $\delta_1$\,=\,$f_\mathrm{aux}$\,--\,$f_\mathrm{n}$ and $\delta_2$\,=\,$f_\mathrm{2n+1}$\,--\,$2f_\mathrm{aux}$, we have $f_\mathrm{ceo}$\,=\,$2f_\mathrm{n}$\,--\,$f_\mathrm{2n}$\,=\,$\mathrm{FSR}$\,--\,($\delta_2$\,+\,$2\delta_1$).

In order to expand the electronic accessing range of the $f_{ceo}$ beatnote, we leverage a down-conversion process as sketched in Fig.\,\ref{figS6}c, where the incident lights in the $f$ and $2f$ paths respectively beat at high-speed photodetectors (PD1 and PD2), and the generated beatnotes are boosted by cascaded low-noise amplifiers (LNA1) before sent into the RF Mixers (Mixer1) for producing down-converted frequency signals below 1\,GHz. After bandpass filtering (BPF1, 20--1000\,MHz), the down-converted beatnotes are boosted by high-gain LNA2s for driving a RF doubler and a Mixer2 in the $f$ and $2f$ paths, respectively. The frequency-doubled signal is selected by another bandpass filter (BPF2) and the mixing frequency signals in the Mixer2 are monitored by an ESA (20 Hz--26.5\,GHz). 
 
The Mixers are driven by two tuned local oscillators (LO1 and LO2) covering a frequency span of 0--40\,GHz ($f_\mathrm{LO1}$) and 0--20\,GHz ($f_\mathrm{LO2}$), respectively. In the experiment, we chose $f_\mathrm{LO1}$ and $f_\mathrm{LO2}$ to be larger than $\delta_1$  and $\delta_2$. As a result, the output frequencies from the Mixer2 read:  
 \begin{equation} 
\begin{split}
   \Delta f_1 = 2f_\mathrm{LO1} - f_\mathrm{LO2} - (2\delta_1 - \delta_2)\\ \Delta f_2 = 2f_\mathrm{LO1} + f_\mathrm{LO2} - (2\delta_1 + \delta_2)
\end{split}
\label{eq4}
\end{equation}
 It is noteable that this scheme allows for an accessible $f$--$2f$ beatnote (that is $2\delta_1$\,+\,$\delta_2$) up to $2f_\mathrm{LO1}$\,+\,$f_\mathrm{LO2}$, which is 100\,GHz in our apparatus.

\def\bibsection{\section*{\textbf{}}}
\large
\section{Supplementary references}